\documentclass[aps,prd,twocolumn,superscriptaddress,showpacs,showkeys,
nofootinbib]{revtex4-1}

\usepackage{latexsym}
\usepackage{amsmath}
\usepackage{amssymb}
\usepackage{amsfonts}
\usepackage{multirow}
\usepackage{color}

\usepackage{supertabular}
\usepackage{placeins}
\usepackage{epsfig}
\usepackage{graphicx}
\usepackage{hyperref}
\hypersetup{colorlinks=true,linkcolor=blue,urlcolor=blue,citecolor=green}

\begin{document}

\title{A chiral quark model analysis of the $\bar KN$ interaction}


\author{M.~Conde-Correa}
\email[]{marlon.conde@epn.edu.ec}
\affiliation{Departamento de F\'isica, Escuela Polit\'ecnica Nacional, Quito 170143, Ecuador.}

\author{T.~Aguilar}
\affiliation{Departamento de F\'isica, Escuela Polit\'ecnica Nacional, Quito 170143, Ecuador.}

\author{A. Capelo-Astudillo}
\affiliation{Departamento de F\'isica, Escuela Polit\'ecnica Nacional, Quito 170143, Ecuador.}

\author{A.~Duenas-Vidal}
\email[]{alvaro.duenas@epn.edu.ec}
\affiliation{Departamento de F\'isica, Escuela Polit\'ecnica Nacional, Quito 170143, Ecuador.}

\author{J.~Segovia}
\email[]{jsegovia@upo.es}
\affiliation{Departamento de Sistemas F\'isicos, Qu\'imicos y Naturales, Universidad Pablo de Olavide, E-41013 Sevilla, Spain.}

\author{P.~G.~Ortega}
\email[]{pgortega@usal.es}
\affiliation{Departamento de Física Fundamental and Instituto Universitario de F\'isica
Fundamental y Matem\'aticas (IUFFyM), Universidad de Salamanca, E-37008 Salamanca, Spain}

\date{\today}
\begin{abstract}
In this work we analyze the $\bar KN$ interaction in the framework of a constituent quark model. The near-threshold elastic and charge exchange cross sections are evaluated, finding a good agreement with the experimental data. Furthermore, the possible existence of $\bar KN$ bound states are explored, finding two poles in the isoscalar $J^P=\frac{1}{2}^-$ sector that can be interpreted as the experimental $\Lambda(1405)$ state.
\end{abstract}


\maketitle



\section{Introduction}

The interest in strangeness in nuclear physics is primarily driven by the distinctive role of the strange quark within low-energy quantum chromodynamics (QCD). Located between the domains of light and heavy quarks, its presence introduces an interaction characterised by spontaneous and explicit chiral symmetry breaking. This breaking pattern gives rise to a remarkably strong attractive interaction between antikaons and nucleons near their respective thresholds, suggesting the possible existence of quasi-bounded states involving antikaons with both nucleons and nuclei, the so-called \emph{kaonic nuclei} (see Refs.~\cite{Iwasaki:2022gwz,Hyodo:2022xhp} for a review).

In fact, the study of the isoscalar $\bar KN$ system lead to the prediction in 1959~\cite{Dalitz:1959dn,Dalitz:1960du}, and latter discovery in 1961~\cite{PhysRevLett.6.698}, of the $\Lambda(1405)$ in the $\pi\Sigma$ invariant mass distribution of the $K^-p \rightarrow \pi\pi\pi\Sigma$ reaction at $1.15$ GeV. This state, with $J^P=\frac{1}{2}^-$~\cite{CLAS:2014tbc}, is compatible with a quasi-bound $\bar KN$ state embedded within the $\pi\Sigma$ continuum with a large decay width of $\sim 50$ MeV, revealing a complex intrinsic quasi-molecular structure. From a quark model point of view, the $\Lambda(1405)$ resonance serves as a pioneering example of an \emph{exotic} baryon, distinguished by its underlying five-quark composition ($udu \bar us$ and $udd \bar ds$).

The discovery of this hyperon-like state, just $27$ MeV below the $K^-p$ threshold, triggered a large number of theoretical and experimental research in order to unveil the nature of the $\Lambda(1405)$, where many authors suggest a two-pole nature in the $\pi\Sigma$ unphysical sheet, one associated to the $\bar KN$ and another to the $\pi\Sigma$ channel (see, e.g., Refs.~\cite{Hyodo:2011ur,Mai:2020ltx,Meissner:2020khl,Tolos:2020aln,Ikeda:2011pi,Hyodo:2022xhp,Ikeda:2012au,Cieply:2011nq,Guo:2012vv,Mai:2014xna,Nieves:2024dcz,Hyodo:2015rnm,Lee:1995ku,Murakami:2023phq,Cid-Mora:2023kuf,Azizi:2023tmw,Miyahara:2015bya}). 

Thus, the study of the interaction of the strange mesons and nucleons embody a crucial aspect of our understanding of exotic hadrons and strange nuclei, with relevance in the study of neutron stars~\cite{Kaplan:1986yq,Tolos:2020aln,Djapo:2008au,RevModPhys.88.021001}.

In this work we analyze the $\bar K N$ system in the framework of a widely used constituent quark model (CQM)~\cite{Vijande:2004he, Segovia:2008zz}, which has been applied to the study of the $N\bar N$ system~\cite{Entem:2006dt}, the $NN$ interaction~\cite{Entem:2000mq} and the deuteron properties~\cite{Valcarce:2005em}. Furthermore, in the last decades it has been successfully employed to describe the phenomenology associated to meson-meson, baryon-meson and baryon-baryon systems~\cite{Ortega:2012rs,Ortega:2022uyu,Ortega:2011zza,Ortega:2012cx}. 
As a result of this careful analysis of the hadron phenomenology, all the parameters of the model have already been constrained.

The paper is organized as follows: After this introduction, Sec.~\ref{sec:model} briefly presents the theoretical framework. In section~\ref{sec:results} the results are analyzed and discussed. Finally, we summarize and draw some conclusions in Sec.~\ref{sec:summary}.

\section{Theoretical Framework}\label{sec:model}

\subsection{Constituent quark model}

For the study of the Antikaon-Nucleon ($\bar KN$) dynamics, with quark content $\bar nsnnn$ where $n=\{u,d\}$, we will use a constituent quark model (CQM) which models the basic phenomenology of Quantum Chromodynamics (QCD) at low and intermediate energies~\cite{Vijande:2004he, Segovia:2008zz, Fernandez:2019ses,Ortega:2022tbl}. 
This CQM is based on the spontaneous breaking of the chiral symmetry at some momentum scale, following the Diakonov's picture of the QCD vacuum~\cite{Diakonov:2002fq} as a dilute instanton liquid.
As a consequence, quarks acquire a dynamical mass due to interactions with fermionic zero modes of individual instantons. This momentum-dependent mass vanishes at high momenta and serves as a natural cutoff for the theory at low momenta. This scenario can be modeled with the following chiral invariant Lagrangian~\cite{Diakonov:2002fq}:
\begin{equation}
\label{ec.1}
    \mathcal{L} = \Bar{\Psi}\left[i\gamma^{\mu}\partial_{\mu} - M(q^2) U^{\gamma_5}\right]\Psi,
\end{equation}
where $U^{\gamma_5} = \exp(i\phi^a \lambda^a \gamma_5 /f_{\pi})$; $\phi^a$ denotes the pseudoscalar fields $\{\vec{\pi}, K_i, \eta_8\}$ with $i = 1 \dots 4$; $\lambda^a$ are the $SU(3)$ flavour matrices; and $M(q^2)$ is the dynamical constituent quark mass. The momentum dependence of the constituent quark mass can be parameterized as $M(q^2)=m_qF(q^2)$, with $m_q\approx 300$ MeV and where 

\begin{equation}
\label{ec.2}
    F(q^2) =\sqrt{\frac{\Lambda^2}{\Lambda^2+q^2}},
\end{equation}
where $\Lambda$ is a cutoff parameter that fixes the chiral symmetry breaking scale.

Expanding the Nambu-Goldstone boson field matrix from the latter Lagrangian we obtain:
\begin{equation}
\label{ec.3}
U^{\gamma_5} = 1 + \frac{i}{f_{\pi}} \gamma_5\lambda^a\phi^a - \frac{1}{2f_{\pi}^2}\phi^a\phi^a + _{\dots}.
\end{equation}

Here, the contribution of the constituent quark mass is identified in the first term. Further terms give rise to quark-quark interactions mediated by boson exchanges. Specifically, the second term represents the exchange of one boson, while the third term illustrates a two-boson exchange, primarily modeled as a scalar $\sigma$ exchange.

\begin{table}[t]
   \caption{ \label{tab.1}Quark-model parameters.}
	\begin{tabular}{lll}
    \hline\hline
		Quark Masses & $m_n$ [MeV] & 313 \\
		                & $m_s$ [MeV] & 555 \\ \hline
	Nambu-Goldstone Bosons &  $m_{\pi}$ [fm$^{-1}$]  & 0.70  \\
		                   &  $m_{\sigma}$ [fm$^{-1}$]  & 3.42  \\
                         &  $\Lambda_{\pi}$ [fm$^{-1}$]  & 4.20  \\
                         &  $\Lambda_{\sigma}$ [fm$^{-1}$]  & 4.20  \\
                         &  $g_{ch}^2/4\pi$  & 0.54  \\
                          &  $\alpha_s$  & 0.497  \\
		\hline\hline
	\end{tabular}
\end{table}

The model is completed with two further QCD effects: the confinement and the one gluon exchange interactions. The first one is a non-perturbative phenomena that prevents from having colorful hadrons, but it does not have a direct contribution to the $\bar KN$ interaction. Regarding the gluon, even below the chiral symmetry breaking scale quarks can still interact via the exchange of one gluon, a QCD perturbative effect which can be described by the Lagrangian~\cite{DeRujula:1975qlm},

\begin{equation}
\label{ec.4}
 \mathcal{L}_{gqq} = i\sqrt{4\pi\alpha_s}\bar{\psi}\gamma_\mu
G_c^\mu\lambda^c\psi,
\end{equation}
being $\lambda^c$ the $SU(3)$ color matrices and $G_c^\mu$ the gluon
field. For the $\bar KN$, direct one-gluon exchanges are not allowed between colorless hadrons, but it will contribute via annihilation diagrams that will be explained below.

The basic non-relativistic potentials at quark level, relevant for the $\bar KN$ system, can be obtained within this model in the static approximation and are given by

\begin{align}
\label{ec.5}
V_{\pi}(\Vec{q}\,) &= - \frac{1}{(2\pi)^3}\frac{g_{ch}^2}{4m_im_j}\frac{\Lambda_{\pi}^2}{\Lambda_{\pi}^2+q^2}\frac{(\Vec{\sigma}_i\cdot\Vec{q})(\Vec{\sigma}_j\cdot\Vec{q})}{m_{\pi}^2+q^2}(\Vec{\tau}_i\cdot\Vec{\tau}_j),\nonumber \\
V_{\sigma}(\Vec{q}\,) &= - \frac{g_{ch}^2}{(2\pi)^3}\frac{\Lambda_{\sigma}^2}{\Lambda_{\sigma}^2+q^2}\frac{1}{m_{\sigma}^2+q^2},
\end{align}
where the $\vec q$ is the transferred momentum, the $\vec \sigma$ ($\vec\tau$) are the Pauli spin (isospin) matrices and $m_{i(j)}$ is the mass of the quark $i(j)$. The parameters of the model, shown in Table~\ref{tab.1}, are constrained by previous studies of hadron phenomenology, e.g., the $NN$ interaction~\cite{Entem:2000mq, Valcarce:2005em}, the $N\bar N$ system~\cite{Entem:2006dt} and other baryon-baryon~\cite{Ortega:2022tbl,Ortega:2011zza,Iglesias-Ferrero:2022zcl} and meson-baryon~\cite{Ortega:2012cx,Ortega:2022uyu} systems involving nucleons and/or strange hadrons.

Two types of interactions are considered in this work, diagrammatically shown in Fig.~\ref{fig:NNb-int}. On the one hand, the exchange of Goldstone bosons between a $K$ meson and a nucleon via the potentials of Eq.~\eqref{ec.5}. On the other hand, the light antiquark of the $K$ meson can annihilate with the quarks inside the nucleon. These are shown in the last two diagrams of Fig.~\ref{fig:NNb-int}. In our model, the real component of this potential can be derived from annihilation diagrams involving the exchange of a gluon or a pion. When represented in momentum space, this interaction can be expressed as~\cite{Entem:2006dt,Faessler:1982qt}:

\begin{align}
\label{ec.6}
	V_{A,\pi}(\vec q\,)&=\frac{1}{(2\pi)^3}\frac{g_{ch}^2}{4m_q^2-m_{\pi}^2}\left( \frac{1}{3} + \frac{1}{2} \vec \lambda_i \cdot \vec \lambda_j \right)\nonumber\\ &
\left( \frac{1}{2} - \frac{1}{2} \vec \sigma_i \cdot \vec \sigma_j \right)
\left( \frac{3}{2} + \frac{1}{2} \vec \tau_i \cdot \vec \tau_j \right),\\
V_{A,g}(\vec{q}\,) &= \frac{\alpha_s}{8\pi^2m_{q}^2}
\left( \frac{4}{9} - \frac{1}{12} \vec \lambda_i \cdot \vec \lambda_j \right)\nonumber\\ &
\left( \frac{3}{2} + \frac{1}{2} \vec \sigma_i \cdot \vec \sigma_j \right)
\left( \frac{1}{2} - \frac{1}{2} \vec \tau_i \cdot \vec \tau_j \right),
\end{align}
the first one ($V_{A,\pi}$) coming from annihilation through a pseudoscalar boson and the second one ($V_{A,g}$) through a gluon.

\subsection{Resonating Group Method}

To extract the interaction between an antikaon ($\bar K$) and a nucleon ($N$) in terms of quark degrees of freedom we make use of the resonating group method (RGM)~\cite{Wheeler:1937zza, Tang:1978zz}. This approach models the $\bar KN$ system as a five-body problem, considering the quark content of the antikaon (one strange quark and one light antiquark) and the nucleon (three light quarks).  The RGM effectively captures the complex quark dynamics within the meson-baryon system, allowing the interaction potential between the antikaon and the nucleon to be decomposed into a direct potential where the natural cutoff is the wave functions of the hadrons.

Hence, the direct kernel is expressed as:
\begin{equation}
\label{ec.7}
 \begin{split}    
     &^{RGM} V_D(\Vec{P}',\Vec{P}) =  \sum_{i\in A, j\in B}\int d\Vec{p}_{\xi'_A} d\Vec{p}_{\xi'_{B1}}d\Vec{p}_{\xi'_{B2}} d\Vec{p}_{\xi_A} \times \\
      &\times d\Vec{p}_{\xi_{B1}}d\Vec{p}_{\xi_{B2}}\, \phi_{A'}^*(\Vec{p}_{\xi'_A}) \phi_{B'}^*(\Vec{p}_{\xi'_{B1}},\Vec{p}_{\xi'_{B2}}) V_{ij}(\Vec{P}',\Vec{P})\times \\
      &\times\phi_A(\Vec{p}_{\xi_A}) \phi_B(\Vec{p}_{\xi_{B1}},\Vec{p}_{\xi_{B2}}),    
 \end{split}
 \end{equation}
where $\vec{P}^{(')}$ is the initial (final) relative momentum of the $\bar KN$, $\Vec{p}_{\xi_{A(B)}}$ are the Jacobi momentum of the meson (baryon) and $V_{ij}$ represents the quark-quark interaction potential within the constituent quark model, where $i (j)$ runs into the constituents of the meson (baryon).

\begin{figure}[t]
\begin{center}
\includegraphics[width=.48\textwidth]{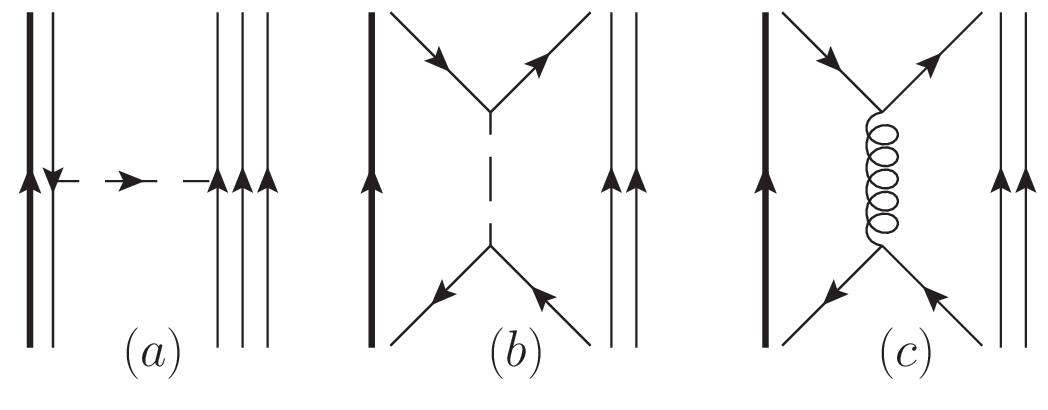}
\caption{\label{fig:NNb-int} Considered interacting diagrams for the $\bar KN$ system: $(a)$ Direct diagrams involving scalar ($\sigma$) or pseudoscalar ($\pi$) Goldstone bosons exchanges between the constituents of the $\bar K$ ($s\bar n$) and the $N$ ($nnn$), $(b)$ Annihilation diagrams through a pion and, $(c)$ Annihilation diagram through a gluon. Thin lines represent light quarks ($n=\{u,d\}$), while thick lines represent a strange quark $s$.  }
\end{center}
\end{figure}

In Eq.~\eqref{ec.7}, $\phi_{A(B)}$ represents the wave function for meson (baryon). On the one hand, the $\bar K$ meson wave function is built as:
\begin{equation}
\label{ec.8}
    \phi_A(\vec q\,) = \Psi_A(\vec q\,) \chi_{\rm ST}^{(A)} \xi_c^{(A)}[1^3],
\end{equation}
where $\chi_{\rm ST}^{(A)}$ is the spin-isospin wave function, $\xi_c^{(A)}$ is the color wave function and $\vec q$ is the relative momentum of the $s\bar n$ system. The momentum wave function $\Psi_A(\vec q\,)$ is obtained by solving the two-body Sch\"odinger equation with the potentials of the constituent quark model, expanded into a sum of Gaussians with ranges in geometrical progression, using the Gaussian Expansion Method (GEM)~\cite{Hiyama:2003cu}. Thus, the internal wave function of the $\bar K$ will be given by

\begin{align}
\Psi_{A} (\vec q\,)=\sum_{n=1}^{n_{\rm max}}N_n\,C_{n} e^{-\frac{q^2}{4\eta_n}},
\end{align}
with $N_{n}=(2\pi\eta_n)^{-3/4}$ and $n_{\rm max}=24$. The $\eta_n$ ranges are taken in geometrical progression, $\eta_n = a_0\cdot a_1^{2(1-n)}$,
which minimizes the number of free parameters to just three, $\{n_{\rm max},a_0,a_1\}$, while ensuring a dense description at short distances~\cite{Hiyama:2018ivm}.

On the other hand, the wave function for the baryon state is similar,
\begin{equation}
\label{ec.9}
    \phi_B = \Psi_B(\Vec{p}_{\xi_\rho},\Vec{p}_{\xi_\lambda})\chi_{\rm ST}^{(B)} \xi_c^{(B)}[1^3],
\end{equation}
with  $\chi_{\rm ST}^{(B)}$ is the totally symmetric spin-isospin wave function and $\xi_c^{(B)}$ is the totally antisymmetric color wave function. The $p_{\xi_\rho}$ is the momentum between two light quarks (called the $\rho$ mode), while the $p_{\xi_\lambda}$ is the momentum between the third quark and the center of mass of the other two light quarks (called the $\lambda$ mode). The internal momentum wave function for the nucleon $\Psi_B$ can be obtained with GEM as it is done for the $\bar K$. However, in Ref.~\cite{Valcarce:1995dm} it was shown, from an analysis of the $nnn$ system in the Born-Oppenheimer approach, that a simpler one-Gaussian function is a good approximation for the long-range regime, 
\begin{equation}
\label{ec.10}
    \Psi_B(\Vec{p}_{\xi_\rho},\Vec{p}_{\xi_\lambda}) = \left[\frac{2b^2}{\pi}\right]^{\frac{3}{4}} e^{-b^2p_{\xi_\rho}^2} \left[\frac{3 b^2}{2\pi}\right]^{\frac{3}{4}} e^{-\frac{3b^2}{4}p_{\xi_\lambda}^2},  
\end{equation}
with $b$ the parameter related to the size of the baryon, fixed to $b=0.518$ fm~\cite{Valcarce:1995dm}.

The direct kernel can be factorized as,
\begin{equation}
\label{ec.11}
    ^{RGM}V_D = 3\sum_{i\in A,j\in B}\mathcal{F}^{(A)}_i \mathcal{F}^{(B)}_j V_{ij},
\end{equation}
where all of them are functions of $\vec Q=\vec P'-\vec P$, the transferred momentum between the $\bar K$ and $N$. The $\mathcal{F}^{(A),(B)}$ are the form factors for the anti-meson and baryon which encodes the information of the hadron wave functions in the $AB\to A'B'$ reaction. A factor $3$ must be added to all diagrams in Fig.~\ref{fig:NNb-int} due to multiplicity. They can be expressed as,

\begin{align}
\label{ec.12}
    \mathcal{F}^{(A)}_i(\vec q\,) =& (4\pi)^{3/2}\sum_{n,n'}^{n_{\rm max}}C_{n}C_{n'}N_{n}N_{n'}\left(\frac{\eta_n\eta^*_{n'}}{\eta_{n}+\eta^*_{n'}}\right)^{3/2}\times\nonumber\\
    &\times e^{-\left(1-\frac{m_i}{m_s+m_n}\right)^2\frac{\vec Q^2}{4(\eta_n+\eta^*_{n'})}}, \\
    \mathcal{F}^{(B)}_j(\vec q\,) =& e^{-\frac{b^2\Vec Q^2}{6}}. 
\end{align}
Here we see that ${\cal F}_A$ relates only to the meson wave function, while ${\cal F}_B$ includes the information of the baryon wave function range. These form factors act as natural cutoffs for the quark-quark potential.

To develop a comprehensive model of the $\bar KN$ interactions, it is imperative to take into account the coupling with other meson-baryon channels and annihilation processes to strange baryons, which are rather intricate. These processes are typically described using microscopic quark-level models such as the $^3P_0$ model~\cite{Aguilar:2024odr} for the coupling to the baryon spectrum or exchange diagrams for the coupling with, e.g., $\eta\Lambda$ or $\pi\Sigma$. In this work we will model the loss of $\bar KN$ flux due to the coupling with nearby channels and the baryon spectrum by means of an optical potential approach, to streamline our calculations and improve model feasibility. This methodological approach has been used previously in the context of the $N\bar N$ interaction~\cite{Entem:2006dt} and the hyperon-antihyperon~\cite{Ortega:2011zza} or the $\Lambda_c\bar\Lambda_c$~\cite{Iglesias-Ferrero:2022zcl} production.

In our study, we adopt a parameterization similar to that used in Ref.~\cite{Entem:2006dt}. This approach allows us to effectively capture the essential dynamics of $\bar KN$ interactions within our modeling framework. By exploiting the optical potential, we aim to provide a robust description of the annihilation processes without the computational complexity associated with full quark-level simulations. This simplified methodology improves our ability to predict and understand $\bar KN$ interactions in different energy ranges, facilitating a deeper understanding of the underlying physics of these interactions. The considered optical potential is then a complex Gaussian model with isospin dependence, given by
\begin{equation}
\label{ec.13}
    V_{opt}^{I}(\Vec{q}) = i\cdot W_i^{I} e^{-b'^2\Vec{q}^2/2},
\end{equation}
where $W_i^I$ and $b'$ are parameters, fitted to experimental near-threshold elastic and charge-exchange cross sections.

\subsection{Solution of the Scattering Problem}

Once we have calculated the meson-baryon effective potential by means of the RGM formulation, we obtain the $T$ matrix from the Lippmann-Schwinger equation in each partial wave, solved using the matrix-inversion method described in Ref.~\cite{Machleidt:1003bo} including the complex optical potential described in Eq.~\eqref{ec.13},

\begin{align}
\label{ec.14}
T^{\alpha'}_{\alpha}(z;p',p) =& V^{\alpha'}_{\alpha}(p',p)+\sum_{\alpha''}\int dp'' p''^{2}\times \nonumber \\
  &  \times V^{\alpha'}_{\alpha''}(p',p'')\frac{1}{z-E_{\alpha''}(p'')}T^{\alpha''}_{\alpha}(z;p'',p),
\end{align}
where $\alpha$ represents the set of quantum numbers for a given partial wave $JLST$, $V$ is the full potential and $E_{\alpha}(q)$ is the non-relativistic energy for the momentum $q$. 

The on-shell $S$-matrix is, then, obtained from the $T$-matrix in the non-relativistic kinematics,
\begin{align}
    S_\alpha^{\alpha'}(E) &= \delta_\alpha^{\alpha'}-2\pi\,i\,\sqrt{\mu_\alpha\mu_{\alpha'}k_\alpha k_{\alpha'}}T_\alpha^{\alpha'}(E;k_{\alpha'},k_\alpha),
\end{align}
with $k_\alpha$ the on-shell momentum of the meson-baryon system.

The $\bar KN\to\bar KN$ elastic and charge-exchange cross sections are given in terms of the scattering matrix elements in each partial wave as,
\begin{align}
\label{ec.15}
    \sigma_{el} &= \frac{\pi}{2p^{2}} \sum_J (2J+1) |1 - S_{el}^{J}|^{2}, \\
    \sigma_{ce} &= \frac{\pi}{2p^{2}} \sum_J (2J+1) | S_{ce}^{J}|^{2},
\end{align}
where $I$ denotes the isospin for the corresponding channel and $p$ is the on-shell relativistic momentum, which improves the phase space description. The $S_{el}$ and $S_{ce}$ terms are combinations of the $S$-matrix in isospin $0$ and $1$ as
\begin{align}
 \label{ec.16}
    S_{el}^{J} &= \frac{1}{2} (S_{J}^{I=1} + S_{J}^{I=0}), \\
    S_{ce}^{J} &= \frac{1}{2} (S_{J}^{I=1} - S_{J}^{I=0}).
\end{align}

The cross section is given as a function of the $\bar K$ momentum in the laboratory reference system, $p_{\rm lab}=p_{\rm cm}\frac{E_{\rm cm}}{m_N}$.

\section{Results} \label{sec:results}

\subsection{Elastic and charge exchange cross section}

\begin{figure}[t]
\begin{center}
\includegraphics[width=.5\textwidth]{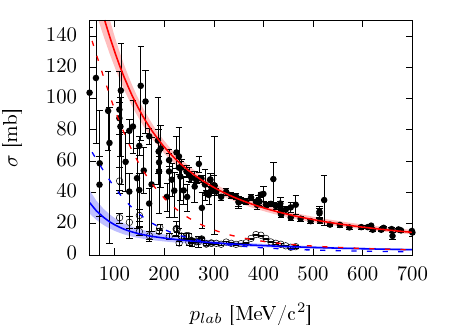}
\caption{\label{fig:xsec} Theoretical $K^-p\to K^-p$ elastic (red) and $K^-p\to \bar K^0n$ cross section (blue), compared to experimental data. Solid lines include the optical potential, the dashed lines are the results without it. Experimental data for the elastic cross section (solid black dots) from Ref.~\cite{ParticleDataGroup:2018ovx} and for the charge exchange cross section (open black dots) from Refs.~\cite{ciborowski1982kaon,kim1966columbia,PhysRevD.14.13,Martin81}. The shadowed band around the theoretical lines show the uncertainty from the fit of the parameters of the annihilation potential. }
\end{center}
\end{figure}

The aim of this work is to study the $\bar KN\to\bar KN$ reactions near threshold. First of all, we analyze the elastic $K^-p\to K^-p$ and the charge-exchange $K^-p\to \bar K^0n$ cross sections below $p_{\rm lab}=700$ MeV/c$^2$, so that the $\eta\Lambda$ channel (with threshold at $\sim 1.66$ GeV) remains closed. In principle, the $\bar KN$ and the $\eta\Lambda$ channels can only be connected by exchange diagrams, which are usually small, so it is safe to ignore the channel. As for the $\pi\Sigma$ channel, its influence is modeled in the optical potential. Another nearby channel is the $\bar K^*N$ (threshold around $1.83$ GeV), which can couple with $\bar KN$ ($\sim 1.43$ GeV), but its influence near the $\bar KN$ threshold was found to be small, so it is not included either. 
We then limit ourselves to the $\bar KN$ channel only. The results for the cross sections with partial waves up to $J=\frac{9}{2}$ are shown in Fig.~\ref{fig:xsec}. We find a good agreement, except for the bump around $p_{\rm lab}\approx 400$ MeV/c$^2$ in the charge exchange cross section due to the $\Lambda(1520)$ baryon, which is not considered in this work. 
 
\begin{table}[t]
\caption{\label{tab:opt} Parameters of the optical potential of Eq.~\eqref{ec.13}, fitted from the experimental elastic and charge-exchange cross sections (see Fig.~\ref{fig:xsec}).}
\begin{tabular}{lc}
 \hline\hline
 $b'$ [fm] & $1.23\pm0.7$\\
 $W_i^0$ [GeV$^{-2}$]& $-0.39 \pm 0.02$ \\
 $W_i^1$ [GeV$^{-2}$] & $-1.28 \pm 0.09$ \\
 \hline\hline
\end{tabular}
\end{table}

For the $\bar KN$ system, no direct $\pi$-exchange is allowed, so the interaction is mainly due to the scalar $\sigma$-exchange and the $\pi$ and gluon annihilation diagrams. The CQM $\bar KN$ interaction alone is capable of describing the charge exchange cross section, but the elastic cross section is smaller than the experimental data, indicating a significant contribution from intermediate states such as baryons or other meson-baryon systems.

The agreement improves when the latter effects are accounted for by the optical potential. The parameters of the optical potential (Eq.~\eqref{ec.13}) are obtained by minimizing the $\chi^2$ function with the available elastic and charge exchange experimental data between $p_{\rm lab}=200$ MeV/c$^2$ and $700$ MeV/c$^2$. In particular, we exclude the region of charge exchange data between $350$ and $450$ MeV/c$^2$, where the $\Lambda(1520)$ resonance signal is prominent. We find a reasonable value of $\chi^2/$d.o.f.$=1.69$ with the parameters of Table~\ref{tab:opt}, where the uncertainty of the optical potential parameters is estimated from the experimental error. 

\subsection{$\bar KN$ molecular states}

Now we analyze the possible existence of $\bar KN$ molecules near threshold. The good agreement of the cross section around the threshold suggests a good description of the $\bar KN$ dynamics in such energy region. Then, it is tempting to explore possible $\bar KN$ bound states in relative $S$-wave.  

The most promising candidate for a $I=0$ $\bar KN$ molecule is the $\Lambda(1405)$, which has been deeply explored since its discovery in 1961~\cite{PhysRevLett.6.698}. A simple baryon picture is unable to reproduce its properties, so a meson-baryon structure must be used.  
In particular, in Ref.~\cite{Fink:1989uk} the meson–baryon scattering amplitude was studied using the bag model of Ref.~\cite{Veit:1984jr}, finding a two-pole structure for the $\Lambda(1405)$. These structures would both contribute to the $\Lambda(1405)$ signal, interfering to form only one resonance. The two-pole structure, emerging from the $\pi\Sigma-\bar KN$ channels, was latter confirmed and analyzed in, e.g., Refs.~\cite{Oller:2000fj,Jido:2003cb,Ikeda:2011pi,Ikeda:2012au,Guo:2012vv,Mai:2014xna}. 

In this work, the effect of the $\pi\Sigma$ channel is encoded in the optical potential, so it is worth exploring if any pole is predicted near the $\bar KN$ threshold. 

First, we analyze the possible structures in $J^P=\frac{1}{2}^-$ without the optical potential, so only with the elastic $\bar KN$ interaction. We include the $^2S_{1/2}$ partial wave in both $I=0$ and $I=1$. We do not find any bound state. However, two virtual states (poles in the second Riemann sheet below the $\bar KN$ threshold) are found for $I=0$ and $I=1$. The $I=0$ has a mass of $1405$ MeV, while the $I=1$ is located at $1414$ MeV. The effect of the $\bar K^*N$ channel is analyzed, including the $^2S_{1/2}-^4D_{1/2}$ partial waves. Its influence is found to be small, though. The $I=0$ pole moves to $1409$ MeV, while the $I=1$ pole moves to $1415$ MeV. 

When we include the optical potential, each virtual state in $I=\{0,1\}$ moves into the complex plane, acquiring width and splitting in two. Then, the isoscalar sector presents two poles, one in $z_1=(1439\pm 3 - i\,22\pm2)$ MeV and another in $z_2=(1417\pm4-i\,55\pm7)$ MeV. In the isovector sector, the two poles are in $z_1=(1444_{-2}^{+3}-i\,7\pm1)$ MeV and in 
$z_2=(1432\pm1-i\,57\pm8)$ MeV.

\begin{table}[t]
\caption{\label{tab:comparison} Position of the poles of $\Lambda(1405)$ (in MeV) found in the $I=0$ second Riemann sheet in this work, compared to other works using chiral SU(3) dynamics.}
\begin{tabular}{lll}
\hline\hline
 Pole 1 & Pole 2 & Reference \\
 \hline
  $(1439\pm 3) - i\,(22\pm2)$ &  $(1417\pm4)-i\,(55\pm7)$ & {\bf This work} \\
 $(1436_{-10}^{+14})-i\,(126_{-28}^{+24})$ & $(1417\pm4)-i\,(24_{-4}^{+7})$ & \cite{Guo:2012vv} \\
 $(1424^{+7}_{-23}) -i\,(26^{+3}_{-14})$ & $(1381_{-6}^{+18})-i\,(81_{-8}^{+19})$ &   \cite{Ikeda:2011pi,Ikeda:2012au} \\
 $1426-16\,i$ & $1390-66\,i$ & \cite{Jido:2003cb}\\
 $(1434\pm2)-i\,(10_{-1}^{+2})$ & $(1330_{-5}^{+4})-i\,(56_{-11}^{+17})$ & \cite{Mai:2014xna}\\
 $1437.7-i\,1.25$ & $1369-i\,71.2$ & \cite{Nieves:2024dcz} \\
 \hline\hline
\end{tabular}
\end{table}

The masses of the two $I=0$ poles are in agreement with other studies performed with chiral SU(3) dynamics, as shown in Table~\ref{tab:comparison}, predicting one wide and one narrower pole around the $\bar KN$ threshold. This result would confirm the two-pole nature of the $\Lambda(1405)$.   

The existence of additional $I=1$ states is also predicted in some previous studies. For example, Ref.~\cite{Oller:2006jw} obtained $I=1$ poles at $1425-i\,6.5$ MeV and $1468-i\,13$ MeV, close to our estimates. In Ref.~\cite{Jido:2003cb}, an $I=1$ state is found, but less stable than the $I=0$ poles. Both states are expected to have a unique resonance structure. However, no experimental state has yet been found in this energy region. This could be due to the fact that these $I=1$ poles are more sensitive to coupled-channels effects than the $I=0$ sector.

\section{Summary} \label{sec:summary}

In this work we have analyzed the $\bar KN$ system in the framework of a constituent quark model where all the parameters are constrained from previous studies of the hadron phenomenology. We have studied the elastic and charge exchange cross section, finding a good agreement with the available experimental data near the $\bar KN$ threshold. 

In addition, we have explored possible bound states in the $J^P=\frac{1}{2}^-$ section, where the $\bar KN$ can be in a relative $S$-wave. If no optical potential is included, we find two virtual states: an isovector state at $\sim 1415$ MeV and an isoscalar state at $\sim 1405$ MeV.

When the effect of other meson-baryon channels and the baryon spectrum is modeled by means of an optical potential, each virtual pole moves into the complex plane and splits in two (see Table~\ref{tab:comparison}), pointing to a two-pole nature for the $\Lambda(1405)$, as suggested by other theoretical works.

\begin{acknowledgments}
This work has been partially funded by Escuela Polit\'ecnica Nacional under projects PIS-22-01, PIS-22-04 and PIM-23-01;
EU Horizon 2020 research and innovation program, STRONG-2020 project, under grant agreement no. 824093;
Ministerio Espa\~nol de Ciencia e Innovaci\'on under grant Nos. PID2019-105439GB-C22, PID2019-107844GB-C22 and PID2022-140440NB-C22;
Junta de Andaluc\'ia under contract Nos. Operativo FEDER Andaluc\'ia 2014-2020 UHU-1264517, P18-FR-5057, PAIDI FQM-370 and PCI+D+i under the title: "Tecnolog\'ias avanzadas para la exploraci\'on del universo y sus componentes" (Code AST22-0001).
\end{acknowledgments}


\bibliographystyle{apsrev4-1}
\bibliography{referencias}

\end{document}